\documentclass[aps,prc,showpacs,twocolumn,nofootinbib,superscriptaddress]{revtex4-1}

\usepackage{longtable}
\usepackage{graphicx}
\usepackage{dcolumn}
\usepackage{bm}
\usepackage{amsmath,amssymb}
\usepackage{longtable}

\newcommand{\ket}[1]{| \, #1 \, \rangle}

\renewcommand\k{{\bm{k}}}
\newcommand\p{{\bm{p}}}
\newcommand\q{{\bm{q}}}

\newcommand\eq{\varepsilon_\q}

\begin{document}


\title{
Universal physics of three bosons with isospin
}


\author{Tetsuo~Hyodo}
\email[]{hyodo@yukawa.kyoto-u.ac.jp}
\affiliation{Department of Physics, Tokyo Institute of Technology, Ookayama, Meguro, Tokyo 152-8551, Japan}
\affiliation{Yukawa Institute for Theoretical Physics, Kyoto University, Kyoto 606-8502, Japan}

\author{Tetsuo~Hatsuda}
\affiliation{Theoretical Research Division, Nishina Center, RIKEN, Wako, Saitama 351-0198, Japan}
\affiliation{Kavli IPMU (WPI), University of Tokyo, Chiba 277-8583, Japan}

\author{Yusuke~Nishida}
\affiliation{Department of Physics, Tokyo Institute of Technology, Ookayama, Meguro, Tokyo 152-8551, Japan}


\date{\today}
\begin{abstract}
We show that there exist two types of universal phenomena for three-boson 
systems with isospin degrees of freedom. In the isospin symmetric limit, 
there is only one universal three-boson bound state with the total isospin 
one, whose binding energy is proportional to that of the two-boson bound 
state. With large isospin symmetry breaking, the standard Efimov states of 
three identical bosons appear at low energies. Both phenomena can be 
realized by three pions with the pion mass appropriately tuned in lattice 
QCD simulations, or by spin-one bosons in cold atom experiments. 
Implication to the in-medium softening of multi-pion states is also 
discussed.
\end{abstract}




\maketitle

\textit{Introduction.}
The properties of particles interacting with a large scattering length are 
universal, i.e., they are determined irrespective of the short range 
behavior of the interaction. In particular, three-particle systems with a 
large two-body scattering length lead to the emergence of the Efimov 
states~\cite{Efimov:1970zz}, which have been extensively studied in cold 
atom physics~\cite{Braaten:2004rn}. Moreover, in condensed matter physics, 
collective excitations in quantum magnets are shown to exhibit the Efimov 
effect~\cite{Nishida:2012hf}.

Since the intrinsic energy scale of the system is not relevant for such 
universal phenomena, they could be also realized in strong interaction 
governed by quantum chromodynamics (QCD) as far as suitable conditions are 
met. In fact, a theoretical possibility of achieving the Efimov effect in 
the three-nucleon system by making a slight modification of the light 
quark masses was discussed in the framework of the effective field theory 
of QCD~\cite{Braaten:2003eu}. The existence of $X(3872)$ near the 
$D^{0}\bar{D}^{*0}$-$\bar{D}^{0}D^{*0}$ threshold has motivated a search 
for the universal three-boson bound state, $D^{0}D^{0}\bar{D}^{*0}$. 
However, the coupled channel effect reduces the attraction and hence no 
universal bound state was found in such a system~\cite{Braaten:2003he}. 

In this Rapid Communication, we show that two universal phenomena arise 
for three bosons with three internal degrees of freedom (such as the pion 
$\pi$) if the two-body scattering length is large. In the real world, the 
$s$-wave $\pi\pi$ scattering lengths are as small as $a_{I=0}\sim -0.31$ 
fm and $a_{I=2}\sim 0.06$ fm~\cite{GarciaMartin:2011cn} due to the 
Nambu-Goldstone nature of the pion, so that the condition of the 
universality does not hold. Indeed, the $I=0$ $s$-wave $\pi\pi$ scattering 
in the real world leads to the $\sigma$ resonance far above the 
threshold~\cite{Beringer:1900zz}. However, if the quark mass is increased 
artificially, the situation changes: the lattice QCD 
simulations~\cite{Kunihiro:2003yj} and chiral effective 
theory~\cite{Hanhart:2008mx} show that the $\sigma$ meson becomes a bound 
state for heavy quark masses. This implies that there is an intermediate 
region of the quark mass where the $\sigma$ meson lies close to the 
$\pi\pi$ threshold and the $\pi\pi$ interaction is characterized by a 
large scattering length. In such a region, the three-pion system exhibits 
universal phenomena, which can be tested by simulating the three pions on 
the lattice by changing the quark mass. From the point of view of the 
statistical noise, three pions with heavy quark mass are much less costly 
than the three nucleons with light quark mass~\cite{Lepage:1989hd}. In 
this sense, the three-pion system is an ideal testing ground for the 
universal physics in QCD.

\textit{Universal physics with the isospin symmetry.}
Let us first consider the three-pion system with exact isospin symmetry. 
We assume that, by an appropriate tuning of the quark mass, only the 
$s$-wave $\pi\pi$ scattering length in the $I=0$ channel, $|a_{I=0}|$, 
becomes much larger than the typical length scale $R$ characterized by the 
interaction range. In addition, we consider momentum $p$ sufficiently 
smaller than $1/R$, so that the pions can be treated as non-relativistic 
particles with a contact interaction. Then, the system is represented by 
the universal zero-range theory~\cite{Braaten:2007nq} with iso-scalar 
interaction: 
\begin{align}
    \mathcal{L}
    = & 
    \sum_{i=1,2,3}
    \phi_{i}^{\dag}\left(
    i\partial_{t}+\frac{\nabla^{2}}{2m}
    \right)
    \phi_{i}
    +v\left|\sum_{i=1,2,3}\phi_{i}\phi_{i}\right|^{2}
    \nonumber \\
    = & 
    \sum_{j=0,\pm}
    \pi_{j}^{\dag}\left(
    i\partial_{t}+\frac{\nabla^{2}}{2m}
    \right)
    \pi_{j}
    +v\left|\pi_{0}\pi_{0}+2\pi_{-}\pi_{+}\right|^{2}
    \label{eq:zerorange} ,
\end{align}
where $\pi_{\pm}=(\phi_{1}\mp i\phi_{2})/\sqrt{2}$ and $\pi_{0}=\phi_{3}$.
Here $m$ and $v$ are the pion mass and the bare coupling constant, 
respectively. The constant interaction is a consequence of the explicit 
symmetry breaking by the nonzero quark mass~\cite{Scherer:2012xha}. It 
dominates over the momentum-dependent interaction near the $\pi\pi$ 
threshold. After the renormalization, the two-pion scattering $t$-matrix 
in $I=0$ is characterized only by the scattering length $a=a_{I=0}$ as
\begin{align}
    it_{0}(p)
    &= \frac{8\pi}{m}
    \frac{i}{\frac{1}{a}-\sqrt{\frac{\p^{2}}{4}-mp_{0}-i0^{+}}}
     \label{eq:twobody} ,
\end{align}
while the $I=2$ component vanishes identically, $it_{2}(p) =0$.

Now we consider the three-pion system. By combining one pion with the 
$\pi\pi$ pair in $I=0$, the whole system has $I=1$, which can also be 
constructed by the $\pi\pi$ pair in $I=2$. Thus, the three-pion system in 
the isospin symmetric limit is a coupled-channel problem. Different 
pairings of two pions are related with each other through the recoupling 
coefficients as
\begin{align}
    &\begin{pmatrix}
      \ket{\pi\otimes[\pi\otimes\pi]_{I=0}}_{I=1} \\
      \ket{\pi\otimes[\pi\otimes\pi]_{I=2}}_{I=1}
    \end{pmatrix}
    = 
    \Omega
    \begin{pmatrix}
      {\ket{[\pi\otimes\pi]_{I=0}\otimes\pi}_{I=1}} \\
      {\ket{[\pi\otimes\pi]_{I=2}\otimes\pi}_{I=1}} 
    \end{pmatrix} ,
    \nonumber \\
    &\quad \quad\quad\quad\Omega = 
    \begin{pmatrix}
    1/3 & \sqrt{5}/3 \\
    \sqrt{5}/3 & 1/6
    \end{pmatrix} .
    \nonumber
\end{align}
Then, the coupled-channel three-body scattering amplitude with total 
energy $E$ obeys the integral equation
\begin{align}
    &iT_{IJ}(E;k,p) \nonumber \\
    &= i\Omega_{IJ}G(P-k-p) \nonumber \\
    &\quad - \sum_{K=0,2}\int_{q} T_{IK}(E;k,q)t_{K}(P-q)
    \Omega_{KJ}G(q)G(P-q-p) ,\nonumber
\end{align}
where $P=(E,\bm{0})$, $k$, and $p$ are the four momenta of the total 
system, of the initial spectator pion, and of the final spectator pion, 
respectively. The pion propagator with momentum $p$ is written as 
$G(p)=1/(p_{0}-\varepsilon_{\p}+i0^{+})$ with $\varepsilon_{\p}
=\p^{2}/2m$.

Because the $I=2$ two-body amplitude vanishes in the present case, the 
coupled-channel three-body equation reduces to a single-channel equation. 
For the on-shell $T$-matrix $T^{\rm on}(E;\k,\p)=T_{00}(E;k,p)|_{k_{0}
=\varepsilon_{\k},p_{0}=\varepsilon_{\p}}$, we have
\begin{align}
    &T^{\rm on}(E;\k,\p) \nonumber\\
    &= \frac{1}{3}g(E,\k,\p) 
    + \frac{1}{3}\int_{\q} T^{\rm on}(E;\k,\q)
    \nonumber \\
    &\quad \times
    g(E,\q,\p)
    \frac{8\pi}{m}\frac{1}{\frac{1}{a} 
    - \sqrt{\frac{\q^2}4-m(E-\eq)-i0^+}} ,
    \label{eq:onshell}
\end{align}
where $g(E,\k,\p)=G(P-k-p)|_{k_{0}=\varepsilon_{\k},p_{0}
=\varepsilon_{\p}}$. Because of the existence of the $I=2$ component, the 
probability of finding an $I=0$ pair after the particle exchange is 
reduced from unity to $1/3$ in the right hand side. This is a major 
difference between identical bosons and the bosons with internal degrees 
of freedom.

\begin{figure}[tbp]
    \centering
    \includegraphics[width=8.6cm,clip]{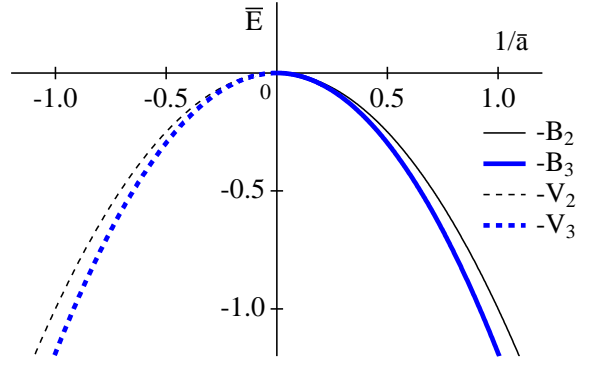}
    \caption{\label{fig:bound}
    (Color online) 
    Energies of the bound and virtual states in the isospin symmetric 
    limit with $\bar{E}={\rm sgn}(E)|mE|^{4}$ and $\bar{a}
    ={\rm sgn}(a)|a|^{4}$ in arbitrary units. The thick (thin) solid line 
    represents the energy of the three-pion (two-pion) bound state, and the 
    thick (thin) dashed line the energy of the three-pion (two-pion) 
    virtual state. 
    }
\end{figure}

Three-pion bound states can be investigated by solving the homogeneous 
equation. After the $s$-wave projection, the scattering amplitude depends 
only on the magnitude of the spectator momenta. At the energy of the bound 
state pole $E=-B_{3}$ with $B_{3}>0$, the amplitude factorizes as 
$T^{\rm on}(E;|\k|,|\p|)=z^{*}(|\bm{k}|)z(|\bm{p}|)/(E+B_{3})$, and the 
eigenvalue equation for $z(|\p|)$ leads to
\begin{align}
    z(|\p|) 
    &= \frac{2}{\lambda\pi}\int_{0}^{\infty}\!d|\q|
    \frac{|\q|}{|\p|}
    \ln\!\left(\frac{\q^2+\p^2+|\q||\p|+mB_3}
    {\q^2+\p^2-|\q||\p|+mB_3}\right) \nonumber \\
    &\quad \times \frac{z(|\q|)}{\sqrt{\frac{3}{4}\q^2+mB_3} 
    - \frac{1}{a}} ,
    \label{eq:boundstate}
\end{align}
where we introduce a factor $\lambda$ for later use; it is $\lambda=3$ in 
the present case. In the two-pion system, there is a bound state at 
$E=-1/(ma^{2})\equiv-B_{2}$ for a positive inverse scattering length 
$1/a$, as seen in Eq.~\eqref{eq:twobody}. This means that the lowest 
threshold in the three-pion channel is $E=0$ (three-body breakup) for 
negative $1/a$ and $E=-B_{2}$ (two-body decay into $\pi$ and a $\pi\pi$ 
bound state) for positive $1/a$. Solving Eq.~\eqref{eq:boundstate} 
numerically, we find one bound state for the three pions at
\begin{align}
    B_{3}
    = \frac{1.04391}{ma^{2}}\quad
    \text{for }1/a>0 .
    \label{eq:bound3}
\end{align}
On the other hand, no bound state is generated for $1/a<0$. Because there 
is no scale other than the scattering length, it is a universal result 
that there is a three-pion bound state whose binding energy is 
proportional to that of the two-pion system. The results are illustrated 
in Fig.~\ref{fig:bound}.

We find that both the three-pion and two-pion bound states vanish at 
$1/a=0$. We can investigate the fate of the bound state pole in the 
negative $1/a$ region. It is known that the two-body bound state pole 
turns into a virtual state, which appears in the lower half-plane of the 
complex momentum space, namely, the second Riemann sheet of the complex 
energy plane~\cite{Taylor}. The first (second) Riemann sheet can be 
reached by a rotation of the binding energy $B_{2}\to B_{2}e^{i\theta}$ 
with $-\pi\leq \theta<\pi$ ($\pi\leq \theta\leq 2\pi$, $-2\pi< \theta
<-\pi$). In fact, when a two-body bound state solution is obtained for a 
given $1/a>0$ at $E=-B_{2}$, there is also a solution for the 
corresponding $-1/a<0$ at $E= -B_{2}e^{2\pi i}=-V_{2}$. Thus, we have a 
two-pion virtual state for a negative $1/a$ at $V_{2}=1/(ma^{2})$. 

We can also search for three-body virtual state solutions by rotating the 
binding energy as $B_{3}\to B_{3}e^{i\theta}$. In general, the 
distribution of the singularities of the integrand of 
Eq.~\eqref{eq:boundstate} depends on the angle $\theta$, so that the 
integration contour should be taken carefully. In the present case, 
however, we can rewrite the equation by using the momentum variables 
$|\tilde{\q}|=|\q|e^{-i\theta/2}$ and $z(|\tilde{\q}|e^{i\theta/2}) =
\tilde{z}(|\tilde{\q}|)$. Noting that the integrand vanishes at 
$|\tilde{\q}|\to \infty$, we obtain the same form as 
Eq.~\eqref{eq:boundstate} with the replacement $1/a\to e^{-i\theta/2}/a$. 
Namely, the rotation of the energy by $e^{i\theta}$ is equivalent to 
rotating the inverse scattering length by $e^{-i\theta/2}$. Thus, as in 
the case of the two-body system, a solution for the virtual state 
($\theta=2\pi$) is obtained by the bound state solution with the opposite 
sign of the scattering length; $V_{3} = 1.04391/(ma^{2})$ for $1/a<0$.

We have numerically checked that there is no resonance state in the second 
Riemann sheet for both positive and negative scattering lengths. In fact, 
the existence of a resonance pole at a certain $\theta$ indicates the 
existence of a pole at $\theta+2\pi$ with the opposite sign of the 
scattering length. However, if $B_{3}e^{i\theta}$ is in the second Riemann 
sheet, $B_{3}e^{i(\theta+2\pi)}$ is in the first Riemann sheet, where the 
existence of poles (except for bound states) is forbidden by 
causality~\cite{Taylor}. Thus, the present system does not allow resonance 
solutions.

\textit{Universal physics with the isospin breaking.}
We now consider the effect of the isospin symmetry breaking. In the real 
world, the charged $\pi_{\pm}$ states are heavier than the neutral 
$\pi_{0}$ mainly due to the electromagnetic interaction. We assume the 
same ordering even for the heavy quark mass and focus on the energy 
window where the effect of $\pi_{\pm}$ can be negligible. Then the system 
reduces to a single-channel problem of three neutral pions. This is 
nothing but the case where the Efimov effect arises~\cite{Braaten:2004rn}.

\begin{figure}[tbp]
    \centering
    \includegraphics[width=8.6cm,clip]{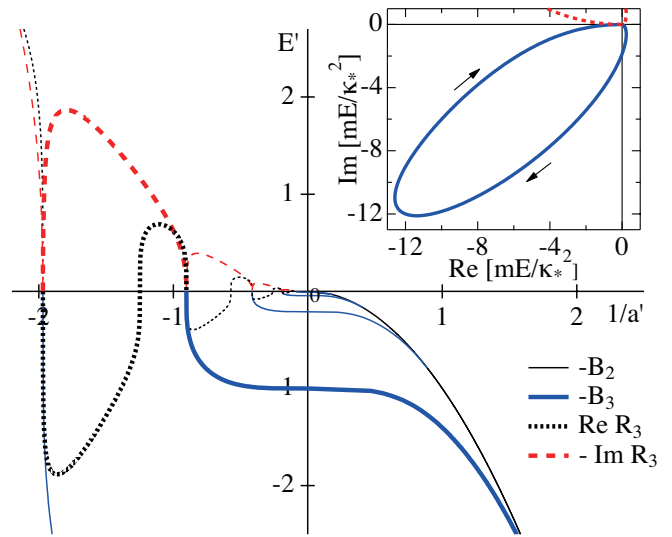}
    \caption{\label{fig:Efimov}
    (Color online) 
    Energies of the bound states of the three $\pi_{0}$ system with the 
    large isospin breaking. The solid lines represent the energies of the 
    bound states, and the dotted (dashed) lines represent the real parts 
    (imaginary parts with the opposite sign) of the resonance poles. Both 
    the axes are normalized to be dimensionless by the Efimov parameter 
    $\kappa_{*}$ as $E^{\prime}={\rm sgn}(E)|mE/\kappa_{*}^{2}|^{1/4}$ and 
    $a^{\prime}={\rm sgn}(a)|a\kappa_{*}|^{1/4}$. The inset shows the 
    trajectories of the pole positions in the complex energy plane for a 
    period corresponding to $-0.663293\geq 1/(a\kappa_{*})\geq -15.0530$. 
    The solid (dashed) line stands for the resonance (anti-resonance) 
    state. As the inverse scattering length is decreased, the pole moves 
    to the direction indicated by the arrows.}
\end{figure}

Let us briefly show how it occurs. Because of the absence of the channel 
coupling, we remove the $1/3$ factor in the right hand side of 
Eq.~\eqref{eq:onshell} and the corresponding eigenvalue equation is 
Eq.~\eqref{eq:boundstate} with $\lambda=1$ as
\begin{align}
    z(|\p|) 
    &= \frac{2}{\pi}\int_{0}^{\infty}\!d|\q|
    \frac{|\q|}{|\p|}
    \ln\!\left(\frac{\q^2+\p^2+|\q||\p|+mB_3}
    {\q^2+\p^2-|\q||\p|+mB_3}\right) \nonumber \\
    &\quad \times \frac{z(|\q|)}{\sqrt{\frac{3}{4}\q^2+mB_3} - \frac{1}{a}}
    f_{\Lambda}(|\q|) ,
    \label{eq:boundstateEfimov}
\end{align}
where $a=a_{\pi_{0}\pi_{0}}$ and $m$ is the mass of $\pi_{0}$. Here we 
introduce the cutoff function $f_{\Lambda}(|\q|)$ with $f_{\Lambda}\to 0$ 
for $|\q|\gg \Lambda$, so that the equation is well defined. For momentum 
much lower than $\Lambda$, the system exhibits the universal physics with 
infinitely many bound states at $1/a=0$. The spectrum in this region is 
independent of the cutoff, and is characterized by the Efimov parameter 
$\kappa_{*}$; the energy of the $n$th excited state is 
$E_{n}\to -e^{-2\pi n/s_{0}}\kappa_{*}^{2}/m$ with $s_{0}=1.00624$ for a 
sufficiently large $n$. In Fig.~\ref{fig:Efimov}, we show by the solid 
lines the energies of the three-pion bound states as functions of $1/a$ 
with appropriate powers for visualization.

To study the trajectory of the bound state pole after the dissociation, we 
search for the resonance solutions in the second Riemann sheet as 
discussed in Refs.~\cite{Yamashita:2002zz,Bringas:2004zz}. For a 
resonance, the residue of the pole should be $z^{*}(|\bm{k}|)z(|\bm{p}|)
\to z(|\bm{k}|)z(|\bm{p}|)$, but the eigenvalue 
equation~\eqref{eq:boundstateEfimov} remains the same. In the present 
case, the rotation of the binding energy $B_{3}\to B_{3}e^{i\theta}$ is 
equivalent to rotating $1/a\to e^{-i\theta/2}/a$ and 
$\Lambda\to \Lambda e^{-i\theta/2}$. In general, the cutoff function can 
have a pole. For $f_{\Lambda}(|\q|)=\Lambda/(|\q|+\Lambda)$, the pole 
position after the phase rotation is $|\tilde{\q}|
=-\Lambda e^{-i\theta/2}$ which does not cross the integration path for 
$-2\pi<\theta<2\pi$ (therefore, our analysis cannot confirm or exclude the 
existence of virtual states). With a different form of the function 
$f_{\Lambda}(|\q|)$, however, we need to treat the singularities properly, 
by adding the residue terms depending on the rotation angle $\theta$, in 
order to obtain the correct results.

By solving the eigenvalue equation, we find that an Efimov bound state 
pole turns into a resonance (and an anti-resonance expressed by the 
complex conjugate of the resonance pole) when the binding energy crosses 
zero for $1/a<0$. The real and imaginary parts of the pole position are 
shown in Fig.~\ref{fig:Efimov}. As we decrease $1/a$, the pole first moves 
to the higher energy with increasing the width, but at some point it 
changes the direction and goes to the subthreshold 
region~\cite{Bringas:2004zz}. Interestingly, as we further decrease $1/a$, 
the pole again moves toward the threshold with decreasing the imaginary 
part, and finally reaches the threshold when the next [$(n-1)$th] Efimov 
bound state comes to the threshold. This pole trajectory is shown in the 
inset of Fig.~\ref{fig:Efimov}. In contrast, when the Efimov bound state 
dissociates with $1/a>0$, the three-body problem is effectively reduced to 
the two-body problem of a pion and a two-pion bound state. According to 
the universal two-body physics, the bound state turns into a virtual 
state, in accordance with Ref.~\cite{Yamashita:2002zz}. In this respect, 
the direct transition from the bound state to the resonance for $1/a<0$ is 
a peculiar feature in the three-body system. Moreover, the cyclic pole 
trajectory is in contrast to the usual two-body resonance which goes away 
from the real axis when the attraction is weakened. This behavior also 
shows that there is only one resonance at a given $1/a<0$, although the 
number of the bound states is infinite at $1/a=0$.

\textit{Classification of the universal physics.}
We have seen that the number of the universal bound states is only one for 
the isospin symmetric case and infinity for the isospin breaking case. The 
difference stems from the factor $\lambda$ in Eq.~\eqref{eq:boundstate} 
which represents the effective reduction of the attraction by the 
coupled-channel effect. The solution of Eq.~\eqref{eq:boundstate} is 
classified by the value of $\lambda$ in Table~\ref{tbl:classification}. 
For $\lambda$ smaller than $4\pi/3\sqrt{3}\approx 2.41840$, the Efimov 
effect occurs. The single universal bound state is obtained for $\lambda$ 
until $3.66811$, above which no bound state appears. The 
$D^{0}D^{0}\bar{D}^{*0}$ three-body system considered in 
Ref.~\cite{Braaten:2003he} corresponds to $\lambda=4$ and hence there is 
no bound state. The three-pion system with heavy quark mass is a unique 
system which is capable of realizing both the Efimov states and the single 
universal state.

\begin{table}[btp]
\caption{Classification of the universal three-body bound states with 
respect to the parameter $\lambda$ in Eq.~\eqref{eq:boundstate}.}
\begin{center}
\begin{ruledtabular}
\begin{tabular}{ccc}
$\lambda<2.41840$ & $2.41840<\lambda<3.66811$ 
& $3.66811<\lambda$ \\
\hline
Efimov states & Single universal state & None \\
\end{tabular}
\end{ruledtabular}
\end{center}
\label{tbl:classification}
\end{table}%

\textit{Interpolation by a model.}
The two types of universal state discussed above can be described 
under the modification of the zero-range theory in 
Eq.~\eqref{eq:zerorange} by introducing a finite mass difference between 
$\pi_{\pm}$ and $\pi_0$, i.e., $\nabla^{2}/(2m)\to \nabla^{2}
/(2m_{j})-m_{j}c^{2}$ with $\Delta=(m_{\pm}-m_{0})c^{2}>0$. The length 
scale associated with the mass difference is $1/\sqrt{m_{0}\Delta}$. If 
this is much larger than the typical interaction range $R$, we should have 
the infinitely many Efimov bound states for $p\ll \sqrt{m_{0}\Delta}$, 
while only a single bound sate arises for $\sqrt{m_{0}\Delta}\ll p 
\ll 1/R$. Measuring the energy from the $\pi_{0}\pi_{0}$ threshold, we 
obtain the two-body amplitude as
\begin{align}
    it(p)
    &= 
    \frac{8\pi}{m_{0}}
    i\Biggl\{\frac{1}{a}
    -\frac{1}{3}\sqrt{\frac{\p^{2}}{4}-m_{0}p_{0}-i0^{+}}
    -\frac{2m_{\pm}}{3m_{0}} \nonumber\\
    &\quad
    \times \left[
    \sqrt{\frac{\p^{2}}{4}-m_{\pm}p_{0}+2m_{\pm}\Delta-i0^{+}}
    -\sqrt{2m_{\pm}\Delta}
    \right]\Biggr\}^{-1} .
    \nonumber
\end{align}
For $\Delta/E\to 0$, we recover Eq.~\eqref{eq:twobody} in the isospin 
symmetric limit if the mass ratio $m_{\pm}/m_{0}$ is close to unity. For 
$\Delta/E\to \infty$, we obtain the $\pi_{0}\pi_{0}$ amplitude with 
$1/a\to 3/a$, and also with an effective range $r_{e}=
-2(m_{\pm}/m_{0})^{2}/\sqrt{2m_{\pm}\Delta}$. Therefore, this model 
interpolates two universal physics with the variation of the energy scale 
$E$. In addition, a cutoff scale for the Efimov physics at 
$\Delta/E\to \infty$ is naturally introduced by the mass difference in the 
form of $r_{e}$. Note that the behavior outside the universal region
should be determined by the underlying dynamics, and this model provides 
one of the possible scenarios.

\begin{figure}[tbp]
    \centering
    \includegraphics[width=8.6cm,clip]{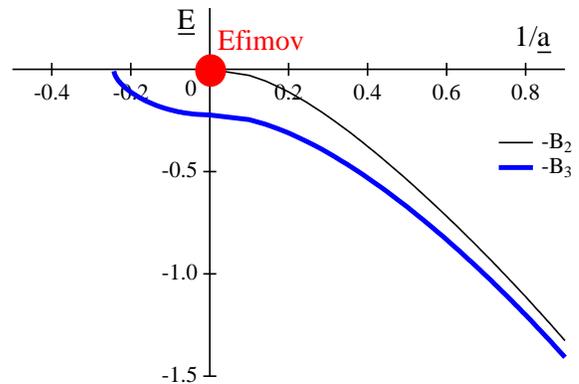}
    \caption{\label{fig:model}
    (Color online) 
    Binding energies from the interpolation model with the mass ratio 
    $m_{\pm}/m_{0}= 1.03403$. The filled circle at origin expresses the 
    accumulation of the Efimov bound states. The axes are given by the 
    dimensionless quantities $\underline{E}={\rm sgn}(E)|E/\Delta|^{1/2}$ 
    and $1/\underline{a}={\rm sgn}(a)|1/a\sqrt{m_{0}\Delta}|^{1/2}$.}
\end{figure}

In Fig.~\ref{fig:model}, we show the spectrum of the above interpolation
model with the mass ratio being $m_{\pm}/m_{0}= 1.03403$. The infinitely 
many Efimov states appear in the vicinity of the origin where the isospin 
breaking scale is much larger than the eigenenergies. For the 
$1/a\gg \sqrt{m_{0}\Delta}$ region, there is only one bound state at 
$B_{3}=1.02982/(m_{0}a^{2})$ which is slightly different from 
Eq.~\eqref{eq:bound3} because of the mass ratio $m_{\pm}/m_{0}\neq 1$. In 
the present model, we find that the lowest level of the Efimov states is 
continuously connected to the single universal bound state in the large 
$1/a$ region.

\textit{Discussions.}
As mentioned in the Introduction, the three-pion system with heavy quark 
mass can be simulated in lattice QCD. First, we need to tune the quark 
mass so that the two pions in the $s$-wave and $I=0$ channel form a 
shallow bound state which we call $\sigma$. Then, we search for the bound 
state in the three-pion system in the $I=1$ channel which we call $\pi^*$. 
For the isospin symmetric case, the ratio of the binding energies between 
$\pi^*$ and $\sigma$ will be precisely $B_{3}/B_{2}=1.04391$ in the 
universal region. The confirmation of the Efimov states which appear under 
the isospin symmetry breaking would be a numerical challenge because the 
binding energies are separated by a large factor, $515$. 

The universal three-pion bound states are unlikely to be realized in the 
vacuum because of the small $\pi\pi$ scattering length for light quark 
masses. On the other hand, in the nuclear medium, one can expect an 
effective increase of the $\pi\pi$ attraction due to partial restoration 
of chiral symmetry~\cite{Hayano:2008vn}, so that the $\sigma$ resonance 
approaches and reaches the $\pi\pi$ threshold before the entire symmetry 
restoration taking place~\cite{Hatsuda:1999kd,Hyodo:2010jp}. Around the 
baryon density where $\sigma$ and $\pi\pi$ are almost degenerate, the 
universal three-pion state $\pi^*$ would simultaneously appear. 
Experimental confirmation of such novel bound states ($\sigma$ and 
$\pi^*$) in the nuclear medium is an interesting problem to be explored in 
hadron-nucleus and nucleus-nucleus reactions.

Finally, we note that our results have direct relevance to cold atom 
physics. In such a case, one can consider a spin-one bosonic atom instead 
of the isospin-one pion. Because the interaction between cold atoms can in 
principle be controlled, it is also interesting to study universal 
three-boson physics with spin degrees of freedom in cold atom experiments.

The authors thank Yoichi Ikeda for helpful discussions. This work was 
partially supported by JSPS KAKENHI Grants No. 24105702, No. 24740152, 
and No. 25887020, by the Yukawa International Program for Quark-Hadron 
Sciences (YIPQS), and by RIKEN iTHES Project.

%
%

\end{document}